\documentclass{aastex631}

\newcommand{\acvn}{$\alpha ^2 \textrm{ CVn }$}
\newcommand{\vtr}[1]{\vec{\mbox{\boldmath$#1$}}}  

\usepackage[utf8]{inputenc}
\usepackage{float}
\usepackage{graphicx}
\usepackage{xmpmulti}
\usepackage{textgreek}

\begin{document}

\title{Modeling the Photometric Variability of Alpha$^2$ CVn with a Dynamical Magnetosphere}

\correspondingauthor{M.\ Virginia McSwain}
\email{mcswain@lehigh.edu}

\author[0000-0002-7305-8321]{Cameron M.\ Pfeffer}
\affiliation{Department of Physics \\
Lehigh University \\
16 Memorial Drive East \\
Bethlehem, PA 18015, USA}

\author[0000-0002-4775-2803]{M.\ Virginia McSwain}
\affiliation{Department of Physics \\
Lehigh University \\
16 Memorial Drive East \\
Bethlehem, PA 18015, USA}

\accepted{July 14, 2022}

\begin{abstract}
    Alpha$^2$ Canum Venaticorum (\acvn) is a strongly magnetic star with peculiar chemical signatures and periodic variability that have been long attributed to the diffusion of magnetic elements through the photosphere, leading to chemical spots across the stellar surface. However, recent studies of other magnetic hot stars are consistent with magnetospheric clouds above the surface. Here we take a renewed approach to modeling \acvn with a simplified dynamical magnetosphere (DM) and a tilted, offset magnetic dipole to reproduce its Transiting Exoplanet Survey Satellite (TESS) variability. Our dipole model also reproduces well the magnetic surface map of \acvn from \citet{silvester_kochukhov_wade_2014}.  Its ultraviolet variability, from IUE archival spectra, is also consistent with traditional reddening models.  We further discuss the implications of a magnetosphere on other observable quantities from the system to conclude that it is unlikely to be present in \acvn.  
\end{abstract}

\section{Introduction}

\acvn has a strong magnetic field on the order of 1 kG \citep{silvester_kochukhov_wade_2014}, which leads to several unusual variability signatures with a period of about 5.5 days \citep{farnsworth1932a}.  The spectral lines of iron, chromium, titanium, and rare earth elements exhibit both intensity and radial velocity variations on this timescale \citep{farnsworth1932a, struve1943}.  The star also exhibits periodic optical variability, with the unusual property that the $B$-band amplitude is smaller than the $V$-band amplitude, causing the star to appear bluer when it is fainter \citep{provin1953}.  
\acvn is not alone in these unusual properties, and it is the prototype of the \acvn class of magnetic and chemically peculiar Ap stars \citep{gcvn2017}.

The optical and UV variability of \acvn stars has long been attributed to ``chemical spots'' across the stellar photosphere.  The premise of this model is that their magnetic fields cause elements to diffuse both vertically and horizontally through the photosphere, leading to nonuniform distributions of various elements \citep{michaud1981}.  These abundance patterns alter the reprocessing of the UV flux \citep{krticka2019}, resulting in an uneven surface temperature distribution.  

On the other hand, hotter O- and early B-type stars with strong magnetic fields have higher mass loss rates due to their radiatively-driven winds, and these outflows are generally confined along the field lines close to the star.  Material that reaches beyond the Kepler corotation radius, $R_\mathrm{K}$, is centrifugally supported against gravity, so it can remain trapped above the photosphere to create a ``centrifugal magnetosphere'' (CM; \citealt{petit2013}).  \citet{townsend_owocki2005} showed that the CM can resemble a rigid, warped, disk-like structure oriented between the rotation and magnetic axes.  Such a CM structure leads to somewhat irregular photometric variations and broad, double-peaked H$\alpha$ emission wings that vary over the rotation period \citep{townsend_etal_2005, petit2013}. Beyond the Alfv\'en radius, $R_\mathrm{A}$, the wind density exceeds the magnetic field energy density, so it may leak out from the outer CM in a centrifugal breakout (CB; \citealt{shultz2020}, \citealt{owocki2020}).  The CB is thought to drive the magnetic reconnection events that accelerate electrons, producing the observed gyrosynchrotron radio luminosities \citep{leto2021, shultz2022, owocki2022}.  Such a CM is generally present in fast rotating, hot magnetic stars with $R_\mathrm{K} < R_\mathrm{A}$ \citep{petit2013}.

For more slowly rotating hot magnetic stars, the magnetically-trapped outflows cannot reach $R_\mathrm{K} > R_\mathrm{A}$, so they fall back onto the star in a ``dynamical magnetosphere'' (DM; \citealt{petit2013}; \citealt{owocki2016}).  For a dipolar magnetic field structure, the cross section of the resulting DM resembles bipolar lobes on either side of the star, which may be asymmetrical if the field is offset from the center of the star.  The DM wraps around the star in a torus shape.  Narrow H$\alpha$ emission, not strongly Doppler shifted by rotation, may partially or strongly fill the central absorption line profile as the circumstellar material confined within the DM passes in front of the star over its rotation period \citep{petit2013}.  The rotating DM also produces relatively sinusoidal photometric variations, possibly including two distinct peaks per rotational cycle, and successfully reproduces the lightcurves of several magnetic O-type stars \citep{munoz2020}.  

In a puzzling twist, evidence for corotating circumstellar clouds in the Bp star HD 37776 was recently published by \citet{krticka2022}.  Although its photometric variability has traditionally been explained by chemical spots similar to those of \acvn \citep{krticka2007}, newly discovered complexity in its TESS lightcurve requires a contribution from corotating clouds in a higher order, multipolar magnetic field.  HD 37776 is not an isolated case, either.  We searched the TESS archives for other Ap stars with similar qualitative behavior and identified several more examples (e.g.\ HD 95974 and HD 123350; \citealt{Bernhard_2015, Hammerich_2016}).
The complex photometric variability of these stars is very similar to the behavior observed in HD 37776, suggesting that the class of \acvn stars may also harbor magnetospheric clouds like their hotter Bp counterparts.

Therefore, in this paper we consider an alternative explanation for the prototype of the class, \acvn.  We apply a simplified DM model to \acvn to explore its potential applicability to this class of stars.
Periodic, low amplitude variability is observed in the TESS lightcurve of \acvn, described in Section \ref{TESSobs}.  In Section \ref{modeling}, we develop a dipole model of the magnetic field which traps circumstellar material, which we use to compare to the measured surface field strength \citep{silvester_kochukhov_wade_2014} and the TESS variability.  We further investigate its UV variability from IUE spectra in Section \ref{iue}.  While the optical and UV properties could be consistent with a DM, we investigate other observables in Section \ref{other} and show that \acvn is probably not a good candidate to possess a magnetosphere.

\section{TESS Observations} \label{TESSobs}

The lightcurve of \acvn used in this paper was produced by the Transiting Exoplanet Survey Satellite (TESS). TESS records lightcurves for over 200,000 of the brightest stars in our vicinity on a wavelength range from 600--1000 nm. Though the primary science mission of TESS is to detect exoplanets through the transit method, the lightcurves recorded by the satellite are useful for characterizing the photometric variability of nearby stars.  

\begin{figure*}
    \centering
    \includegraphics[scale = .35]{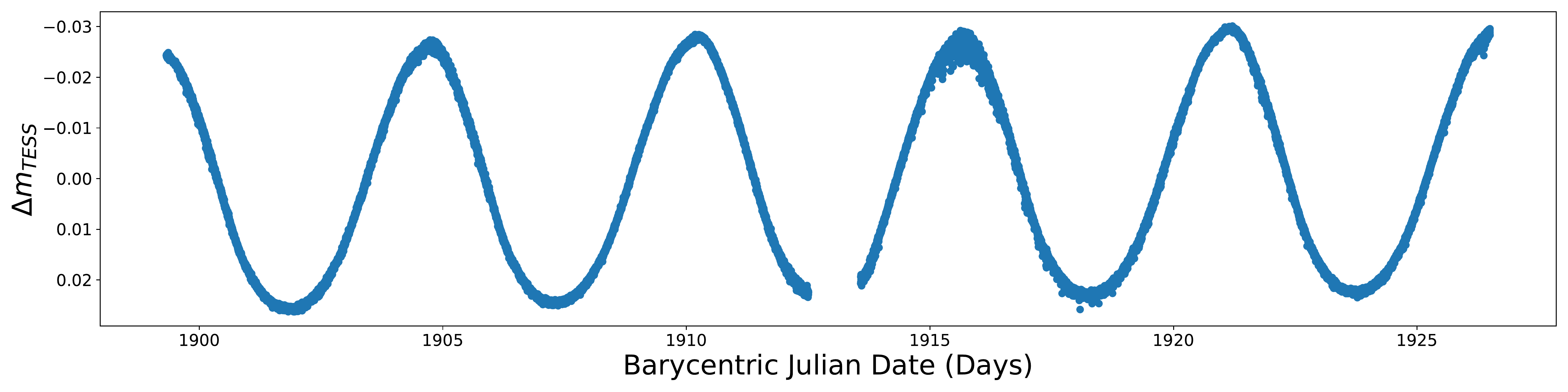}
    \caption{The TESS lightcurve for \acvn showing photometric variations over 27.17 days. The lightcurve exhibits periodic variability with a period of 5.46939 days.}
    \label{TESSLightcurve}
\end{figure*}

 We used the \texttt{astroquery.mast} routine to retrieve the pipeline-reduced TESS data products for \acvn from the Barbara A. Mikulski Archive for Space Telescopes (MAST).  The \texttt{TESSLightCurveFile} object includes both the minimally-processed Single Aperture Photometry (SAP) flux and the Pre-search Data Conditioning SAP flux (PDCSAP) flux that attempts to remove long term trends.  While the PDCSAP flux usually has fewer systematic trends than the SAP flux, in this case the extreme brightness of \acvn ($V = 2.88$) means that the SAP flux provides a significantly cleaner measurement of its relative flux variations.  With such a bright star, the absolute SAP flux measurements may not be very accurate, but we are more concerned with the differential variability rather than the absolute flux in this work.  As we show below, the period and small amplitude of the SAP flux variations are highly consistent with prior studies of \acvn, so the default pipeline product is sufficient for our needs. We computed the mean value of the SAP flux and used the logarithmic ratio to produce the differential lightcurve shown in Figure \ref{TESSLightcurve}.
 
The period of the variations in the lightcurve is interpreted as the rotation period of the star \citep{petit_1987}.  We used the \texttt{LombScargle} routine in the \texttt{astropy.timeseries} package \citep{astropy2018} to measure the rotation frequency,  
which we find to be 0.1828 d$^{-1}$.  This corresponds to a period of 5.47(1) d, very close to the value of 5.46939 d found by \citet{farnsworth1932b}.  In fact, folding the TESS data on their period resulted in no discernible difference, so we adopt their value for $P$ in this work.  We find that the epoch of maximum brightness, $T_0$, is $BJD - 2400000 = 51921.210$. 

We binned the folded flux into 100 bins for further inspection.  
The standard deviation in each bin provides the size of its error, $\sigma_{obs}$, for the folded lightcurve and our fitting procedure below.

\section{Modeling the Magnetospheric Clouds} \label{modeling}

\subsection{Dipolar Magnetic Field Configuration}

We adopt a stellar radius of 2.49 $R_\odot$ and inclination of  $i = 120^\circ$ \citep{kochukhov_wade_2010} for \acvn. We define the $z$ axis along the rotation axis such the $x,y$ plane at $z = 0$ defines the equatorial plane. 
We model the magnetic field of \acvn as an off-center dipole, tilted relative to the $z$ axis:
\begin{equation}
\vec B(\vec r, \vec m) = {{3*\vec r*(\vec m \cdot \vec r) }\over{{\vec r} ^5}} - {{\vec m} \over { {\vec r}^3}}
\end{equation}
\citep{griffiths_2014}.
We consider 4 parameters when generating the magnetic field: the relative offsets $x_0$, $y_0$, and $z_0$ of the field with respect to the center of the star (in units of $R_\star$) and the obliquity angle $\beta$ between the the $z$ axis and the magnetization axis of the dipole, $\vec m$.  We assume a unit dipole for $\vec m$, which we scale later according to the observed differential magnitudes.  

Our dipole model of \acvn is very similar to the measured surface magnetic field found with Stokes IQUV Magnetic Doppler Imaging \citep{silvester_kochukhov_wade_2014}.  
We compare the surface $B$ field strength predicted by our model and the observed surface map by varying the $x_{\mathrm{0}}$, $y_{\mathrm{0}}$, and $z_{\mathrm{0}}$ offsets from $-0.4 R_\star$ to $0.4 R_\star$ using a step size of 0.1 $R_\star$, varying $\beta$ from 0 to $\pi$ in increments of $\pi/10$. We also tried varying the magnitude of the dipole moment, but we found that the observed magnetic map deviates from a simple dipole too much to constrain this parameter beyond the order of magnitude of $|\vec{B}| \sim 0.8$ kG.  The error in the magnetic map is about 15\% of the maximum value of $B$ (Kochukhov, private communication), so we adopted the value $\sigma_{\mathrm{obs, map}} = 0.682$ kG.  
We defined the error in our map fit using $\sigma_{\mathrm{map}} = \sqrt{  \sum \left( \frac{ (O-C)}{\sigma_{\mathrm{obs, map}}} \right )^2 \frac{1}{ N_{\mathrm{map}}} }$.  The number of points in the surface map is $N_{\mathrm{map}} = 695$.

We find that the map fits offer reasonably good constraints for $x_\mathrm{0}$, $z_\mathrm{0}$, and $\beta$; however the map did not constrain $y_\mathrm{0}$ well.  
Overall, we find a good match to the bulk properties of the radial, azimuthal, and meridional components of the surface field of \acvn. The TESS lightcurve offers further constraints on the field, which we discuss in the next subsection.

\begin{deluxetable}{ccccccc}
\tablecaption{Results from Fitting \label{tab:rms}}
\tablewidth{0pt}
\tablehead{
\colhead{$x_0$} & \colhead{$y_\mathrm{0}$} & \colhead{$z_\mathrm{0}$} & \colhead{$\beta$} &
 \colhead{Lightcurve Fit}  & \colhead{k} & \colhead{Map Fit}  \\
\colhead{($R_\star$)} & \colhead{($R_\star$)} & \colhead{($R_\star$)} & \colhead{(rad)} & \colhead{Error ($\sigma_{\mathrm{lc}}$)} & \colhead{ }  & \colhead{Error ($\sigma_{\mathrm{map}}$)} 
}
\decimalcolnumbers
\startdata
0.0&0.3&-0.3&6$\frac{\pi}{10}$&0.192&0.349&3.294\\
0.0&-0.3&-0.3&4$\frac{\pi}{10}$&0.192&0.349&3.173\\
0.0&0.3&-0.4&6$\frac{\pi}{10}$&0.169&0.38&3.796\\
0.0&-0.3&-0.4&4$\frac{\pi}{10}$&0.169&0.38&3.64\\
0.0&0.3&-0.2&6$\frac{\pi}{10}$&0.214&0.45&3.035\\
0.0&-0.3&-0.2&4$\frac{\pi}{10}$&0.214&0.45&2.943\\
0.0&0.3&-0.1&6$\frac{\pi}{10}$&0.235&0.594&2.903\\
0.0&-0.3&-0.1&4$\frac{\pi}{10}$&0.235&0.594&2.845\\
0.0&0.3&0.0&6$\frac{\pi}{10}$&0.254&0.597&2.857\\
0.0&-0.3&0.0&4$\frac{\pi}{10}$&0.254&0.597&2.836\\
\enddata
\tablecomments{Table \ref{tab:rms} lists the parameters and errors for each lightcurve and surface map model of \acvn. The first 10 lines are shown here as an example, and the full table is available in the electronic version of this paper. Columns 1, 2, and 3 show the offsets of the dipole from the center of the star. Column 4 shows the obliquity angle $\beta$. Column 5 show the errors in the lightcurve fit, $\sigma_{\mathrm{lc}}$, and Column 6 shows the scaling factor $k$ used to scale the model to the TESS lightcurve.  Column 7 contains the errors in the map fit,  $\sigma_{\mathrm{map}}$.}
\end{deluxetable}

\subsection{Lightcurve Fitting}

Following the arguments of \citet{munoz2020}, we model the density of circumstellar particles to be proportional to the magnetic field strength.  Due to the very low mass loss rates of \acvn stars ($\sim 10^{-12} \; M_\odot$~yr$^{-1}$; \citealt{leto2006}) and the relatively small amplitude of the variability compared to other classes of variable stars ($\Delta m_{\mathrm{TESS}} \sim 0.05$ mag; see Fig.\ \ref{TESSLightcurve}), we assume that any putative magnetospheric cloud is optically thin.  
The differential magnitude $\Delta m$ is thus related to the optical depth $\tau$ according to 
\begin{equation}
\Delta m = -2.5\tau\log_{10}(e) = 1.086\tau.
\end{equation}
In turn, $\tau$ is determined by the absorption cross section $\sigma$ (which we assume constant) and the number density of particles $n$ integrated along the line of sight $dl$,
\begin{equation}
\tau = \int \sigma \; n \; dl = k \int B \; dl.
\end{equation}
The scaling factor $k$ relates the proportionality between $B$, $n$, and $\tau$.

\begin{figure} 
    \centering
    \includegraphics[scale = .25]{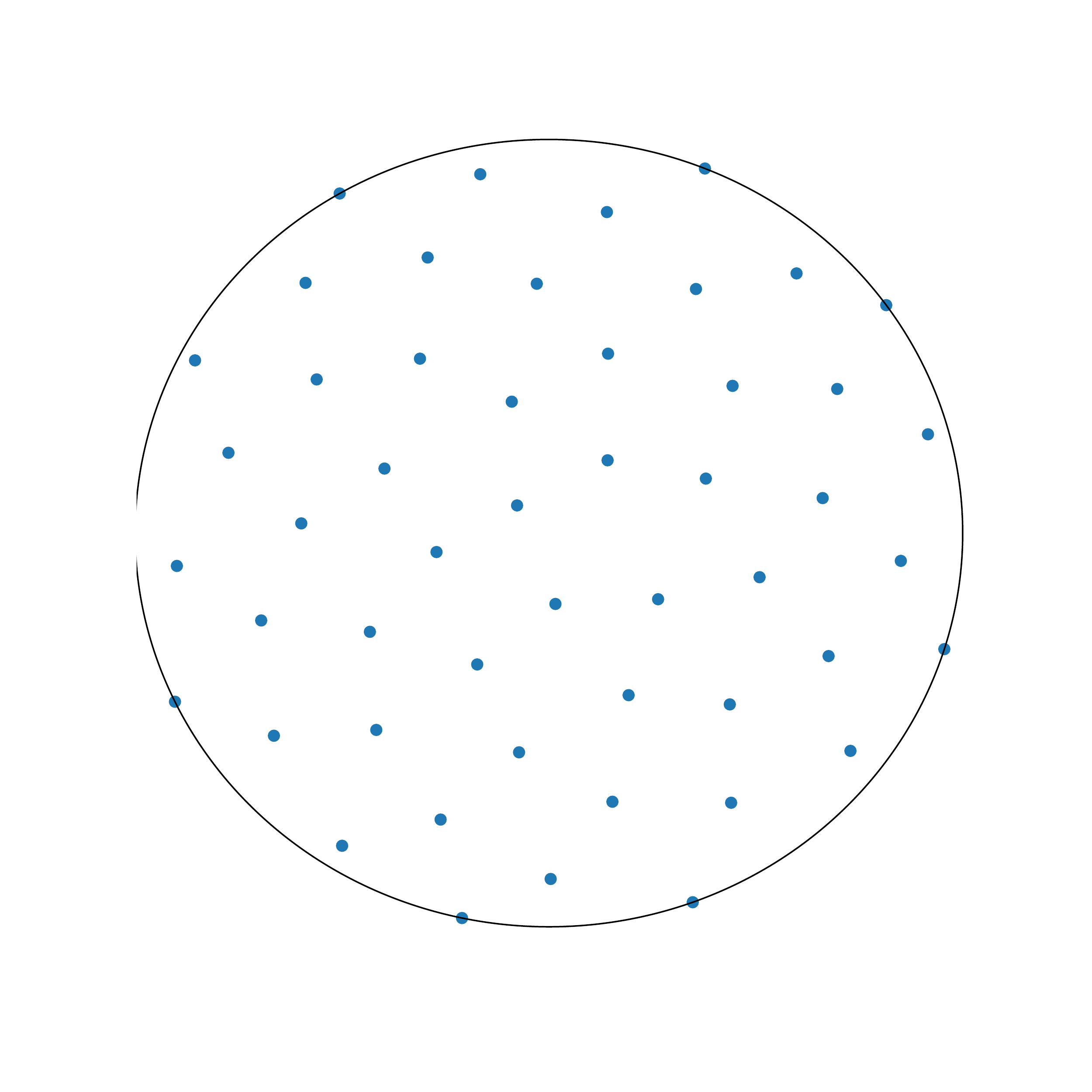} 
    \caption{The blue points, distributed using a ``sunflower seed'' pattern, represent the center of each cylindrical column used to integrate over the dipolar field over each line of sight as the star rotates.  
    The surface of the star is represented with the large black circle.}
    \label{sunflower}
\end{figure}

We perform the numerical integration by summing up the strength of a unit dipolar field at every point along a cylindrical column projected against the star along the observer's line of sight, starting at the stellar surface and extending out to a distance $5 R_\star$, using a step size of $0.25 R_\star$. Since the dipole falls off in strength with distance $r$ according to $1/r^3$, the strength of $\vec B$ falls to less than 1/100$^{th}$ of its surface value and is thus neglected at larger distances.  To fill in the cross sectional area of the column, we use a ``sunflower seed'' distribution \citep{vogel_1979} with 50 points, thus having an average separation of about $0.246 R_\star$.  This separation represents the average thickness of each column.  The distribution of columns is illustrated in Figure \ref{sunflower} and remained the same for every line of sight.  The sunflower seed distribution offers the advantage of a more uniform spatial density of points than a truly random distribution, but without introducing biases from a purely symmetrical distribution.  We tested our model using a denser grid of points, with average separations as low as $0.05 R_\star$, and we found no significant changes in the lightcurve properties.

\begin{figure*} 
    \centering
    \includegraphics[scale = .5]{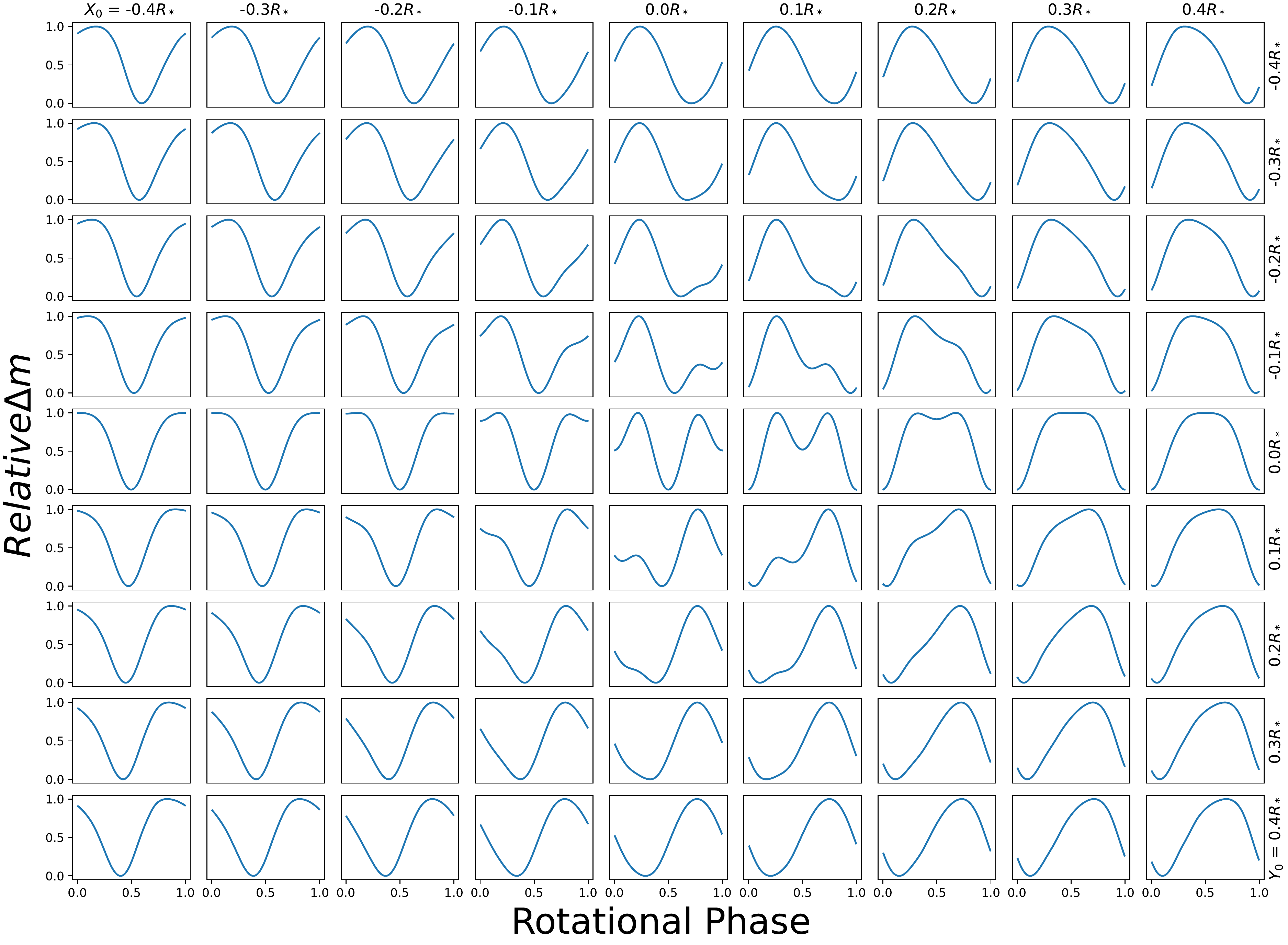}
    \caption{Family of lightcurves for a range of values in $x_{\mathrm{0}}$ and $y_{\mathrm{0}}$.  Here, the parameters $z_{\mathrm{0}}$ and $\beta$ are fixed at $0R_\star$ and $4 \pi/10$, respectively. The amplitude of each lightcurve is arbitrary as the amplitude is scaled linearly to match the TESS amplitude during the fitting procedure.}
    \label{family1}
\end{figure*}

We generate an array of model lightcurves by varying the $x_{\mathrm{0}}$, $y_{\mathrm{0}}$, and $z_{\mathrm{0}}$ offsets from $-0.4 R_\star$ to $0.4 R_\star$ using a step size of 0.1 $R_\star$ and varying $\beta$ from 0 to $\pi$ in increments of $\pi/10$.  Some representative lightcurves are illustrated in Figure \ref{family1}.  The full list of parameters $x_{\mathrm{0}}$, $y_{\mathrm{0}}$, $z_{\mathrm{0}}$, $\beta$, and $k$ values are listed in Table \ref{tab:rms}.  Overall, when we compared our (non-offset) model lightcurves to those of \citet{munoz2020}, we found excellent agreement.  This implies very small contributions of the wind upflow and the hot post-shock gas in their models, which supports our decision to neglect those components.

We calculate the integrated sum $\int B \; dl$ along each column for 100 different lines of sight to represent each rotational phase.  To fit these model lightcurves to the TESS lightcurve of \acvn,  we looped over every model, scaling them vertically with the factor $k$ to match the amplitude of the TESS lightcurve.
To match the phase of the TESS lightcurve, we also applied a cross correlation to the model lightcurve.  The cross correlation has the effect of collapsing the dependence on both $x_\mathrm{0}$ and $y_\mathrm{0}$, so that only one horizontal offset really seems to matter in the model.
We then calculated the residual, $O - C$, for each phase bin and measured its deviation from the observed lightcurve in units of $\sigma_{\mathrm{obs, lc}}$.  We tabulated the total deviation of each model using $\sigma_{\mathrm{lc}} = \sqrt{\sum \left ( \frac{ (O-C)}{\sigma_{\mathrm{obs, lc}}} \right )^2  \frac{1}{N_{\mathrm{lc}}} }$ to find the best fit, where $N_{\mathrm{lc}} = 100$.

The lightcurve constrains $x_{\mathrm{0}}$ well, with the best fit having the values $x_{\mathrm{0}} = 0^{+.1}_{-.1}R_\star$. However, the lightcurve did not offer good constraints on the value of $z_{\mathrm{0}}$.  We found equally good fits with $y_{\mathrm{0}} = .3R_\star$ and $\beta = 6\pi/10$ as with the solution $y_{\mathrm{0}} = -.3R_\star$ and $\beta = 4\pi/10$.  These degenerate solutions are symmetric about the equatorial plane of the star.  From our constraints on $\beta$, we find that the dipole lays nearly perpendicular to the rotation axis, tilted $\approx \pm \pi/10$ relative to the equator.

\begin{figure*}
    \centering
    \includegraphics[scale = .24]{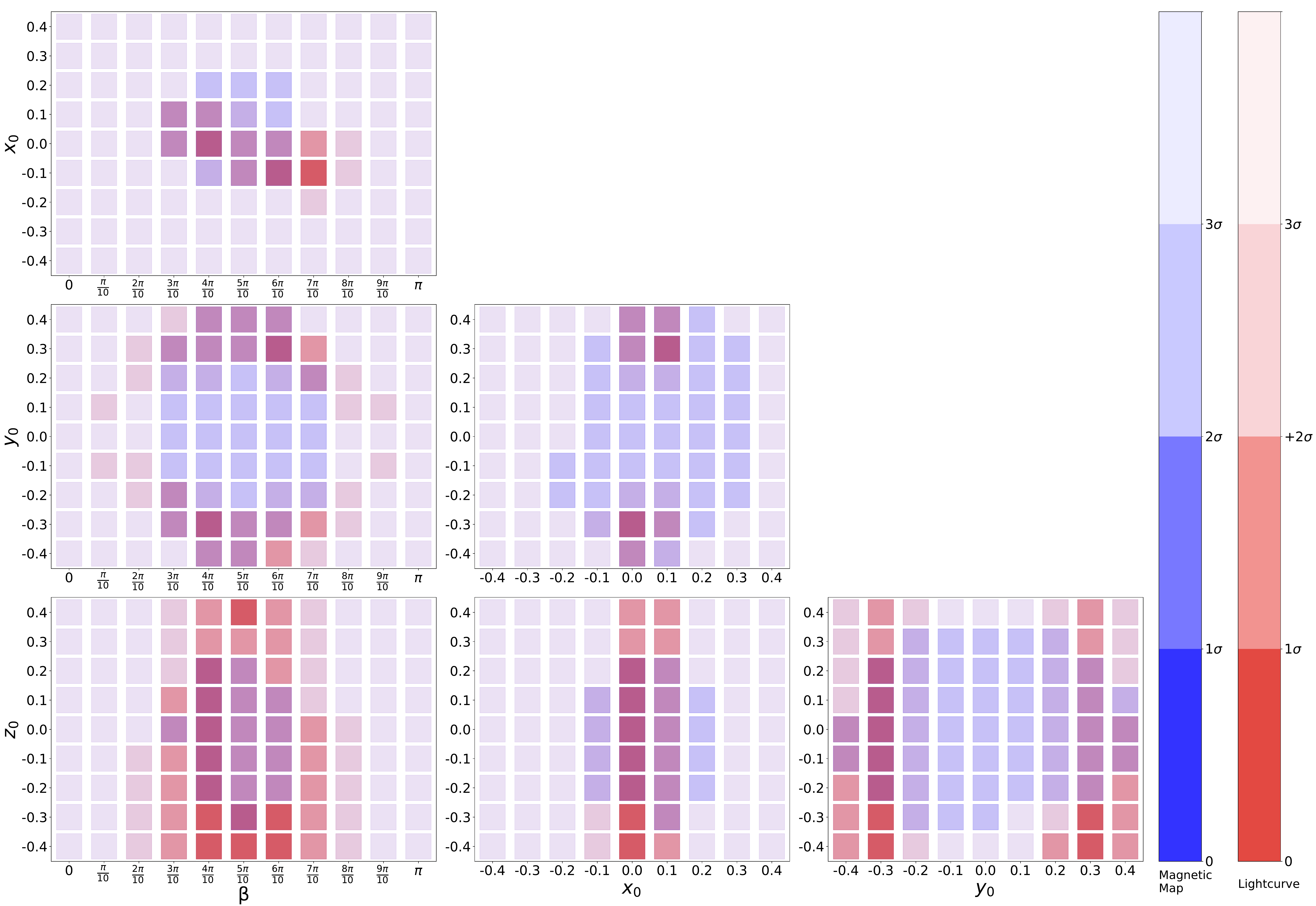}
    \caption{Shown here are the overlapping error grids for the lightcurve and magnetic map models. The grid is centered on the parameters $(x_{\mathrm{0}},y_{\mathrm{0}},z_{\mathrm{0}},\beta) = (0R_\star, -.3R_\star, 0R_\star, 4 \pi/10)$.  
    The errors from our magnetic map fits are shown in shades of blue, and the errors from our lightcurve fits are shown in shades of red.  The purple regions represent parameters that fit both models simultaneously. 
    }
    \label{OverlapRMSGrid}
\end{figure*}

\begin{figure}
    \centering
    \includegraphics[scale=0.4]{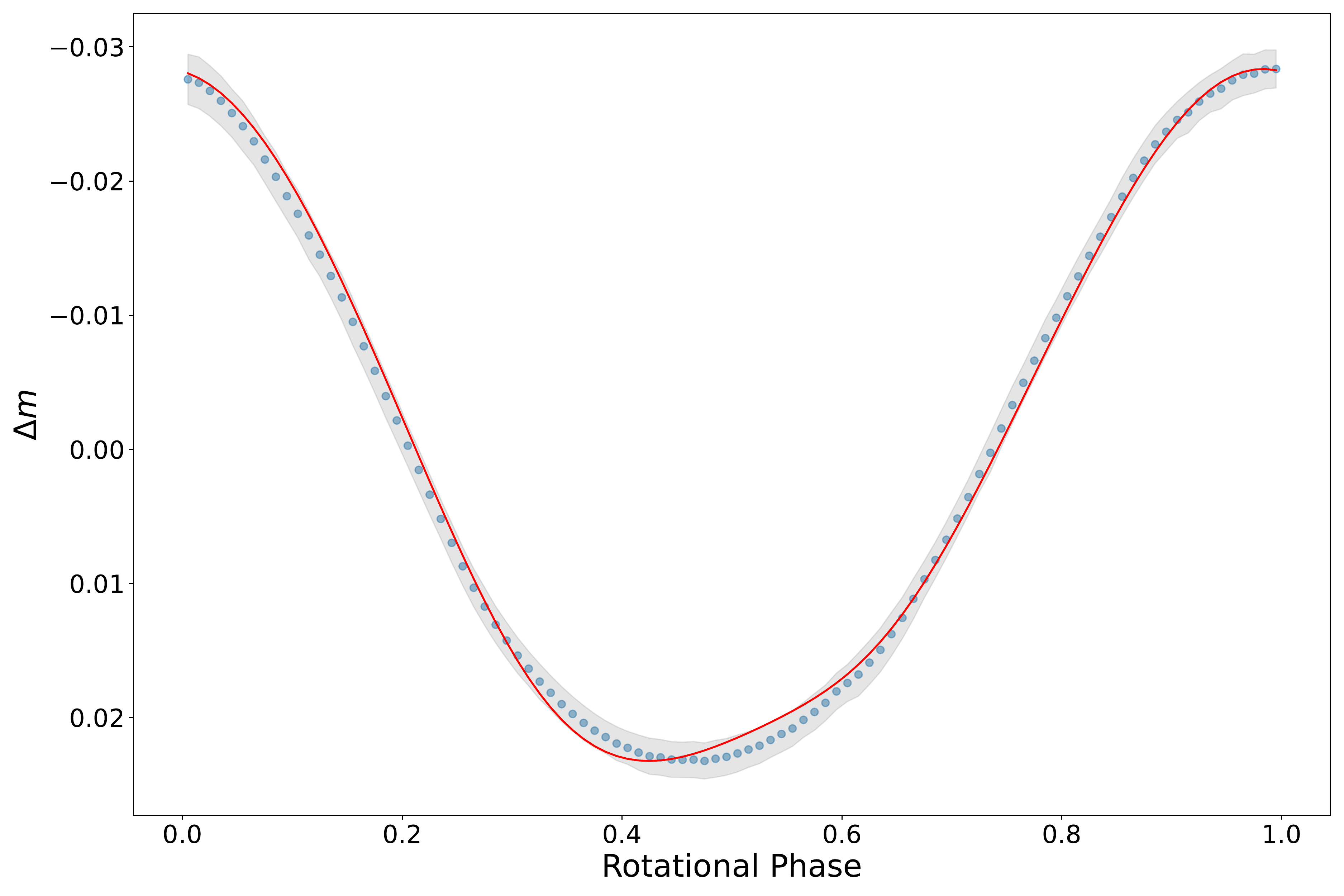}
    \caption{Our adopted model lightcurve for \acvn with parameters  $(x_{\mathrm{0}},y_{\mathrm{0}},z_{\mathrm{0}},\beta) = (0R_\star, -.3R_\star, 0R_\star, \frac{4 \pi}{10})$ is shown here as a red line, and the binned TESS lightcurve is shown as blue points.  The shaded gray region illustrates $\sigma_{\mathrm{obs, lc}}$.
    }
    \label{BoBWLightcurve}

\end{figure}

\begin{figure*}
    \centering
    \includegraphics[scale  = .33]{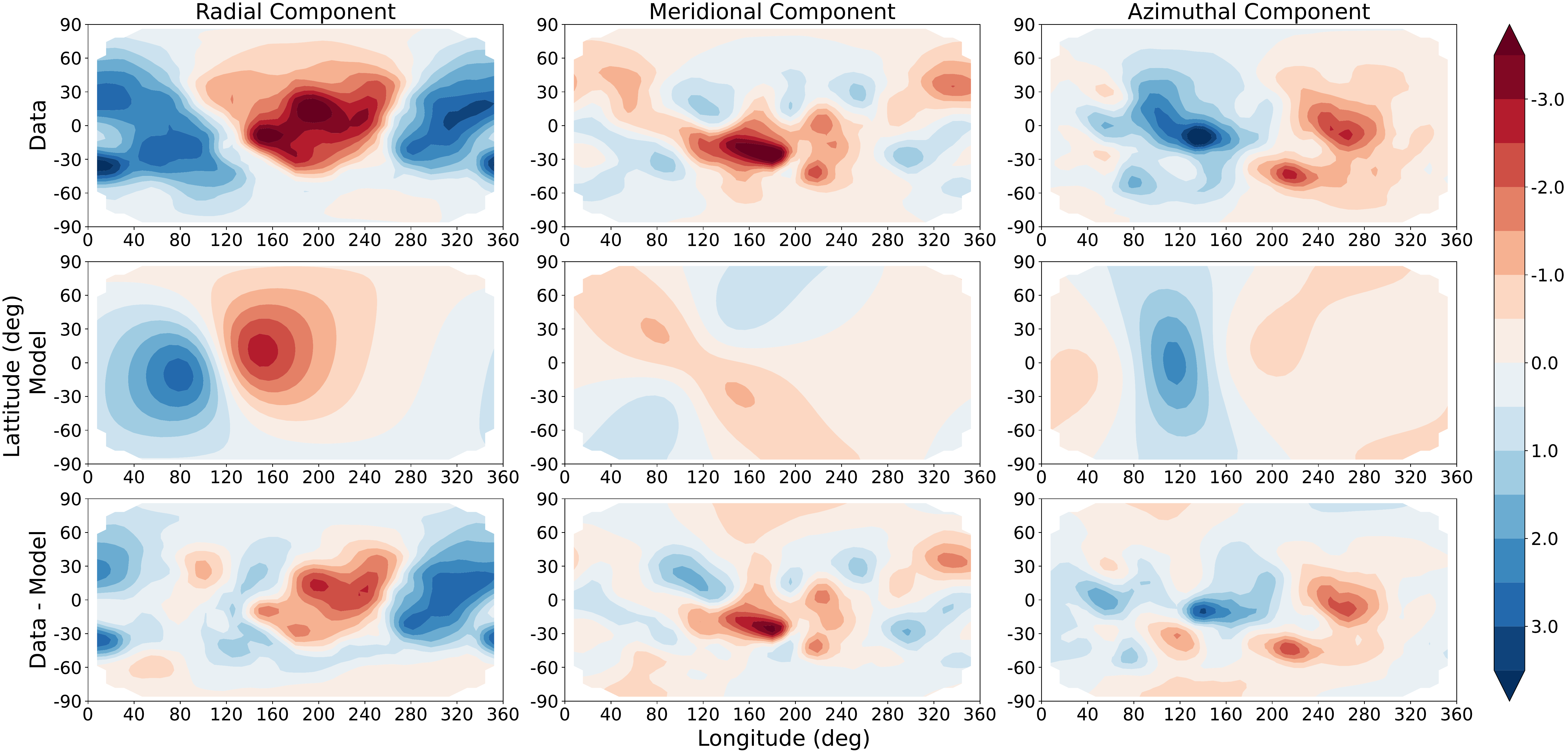}
    \caption{The first row shows the radial (left), meridional (center), and azimuthal (right) components of \acvn's surface magnetic field as a function of stellar latitude and longitude as measured using Stokes IQUV Magnetic Doppler Imaging  \citep{silvester_kochukhov_wade_2014}.  The second row shows the components of our best fit dipole model, shown in the same format, with parameters $(x_{\mathrm{0}},y_{\mathrm{0}},z_{\mathrm{0}},\beta) = (0R_\star, -.3R_\star, 0R_\star, \frac{4 \pi}{10})$.  The third row shows the residuals from our fit.  The colorbar on the right side shows the magnetic field strength in units of kG for each panel. 
    }
    \label{BoBWMap}
\end{figure*}

From the simultaneous constraints offered by the surface map and lightcurve fitting, we combined these results to determine a final solution to our dipole model. The overlapping error grid is shown in Figure \ref{OverlapRMSGrid}. Through this method, we adopt the best fitting parameters as follows: $x_{\mathrm{0}} = 0^{+.1}_{-.1}R_\star$, $y_{\mathrm{0}} = -0.3^{+.1}_{-.1} R_\star$, $z_{\mathrm{0}} = 0^{+.2}_{-.2}R_\star$, $\beta = 4 \pi/10^{+\pi/10}_{-\pi/10}$.  We note that the solution with $y_{\mathrm{0}} = +0.3 R_\star$ and $\beta = 6 \pi /10$ produces an equally valid solution given the symmetry of the dipole model.  The model lightcurve with our adopted best fit is shown in Figure \ref{BoBWLightcurve}, and the corresponding model magnetic map is shown in Figure \ref{BoBWMap}.

We used our opacity model to calculate the total mass of the absorbing particles in the dynamical magnetosphere of \acvn.  Using a standard reddening model (e.g. \citealt{fitzpatrick2004}, hereafter F04), the extinction at the TESS effective wavelength of 7865 \AA, relative to the $V$-band, is $A_{\mathrm{TESS}} / A_\mathrm{V} \approx 0.6$.  Using the relation between neutral hydrogen column density and optical extinction $N_\mathrm{H} / A_\mathrm{V} = 2.21 \times 10^{21}$ atoms~cm$^{-2}$ relation (\citealt{guver2009} and references therein), we find a total mass of absorbing particles of $4 \times 10^{-11}$ $M_\odot$, or an average particle density of about $8\times 10^{11}$ particles cm$^{-3}$.  This compares well to the solar corona, where the B field is about 3 orders of magnitude smaller and the particle density is about $10^8$ to $10^9$ cm$^{-3}$ \citep{sakurai2017}. We also find a maximum particle density of ${8} \times 10^{13}$ particles cm$^{-3}$, and a minimum particle density of $4 \times 10^{10}$ particles cm$^{-3}$.

\section{UV Variability} \label{iue}

The optical properties of \acvn are very consistent with the simplified DM model, and here we expand our model to the UV. 
We use the flux-calibrated, low-dispersion spectra from the \textit{International Ultraviolet Explorer} mission archives to probe the UV flux variability of \acvn.  The archive contains 11 spectra obtained with the Short Wavelength Prime (SWP) camera, 3 with the Long Wavelength Prime (LWP), and 4 with the Long Wavelength Redundant (LWR) cameras.  We use a Gaussian smoothing kernal with standard deviation of 3 pixels to reduce the noise, and we constructed artificial Gaussian bandpasses across 13 regions of the spectra using the same method as \citet{krticka2019}.  From the logarithm of the enclosed flux, we measured the apparent magnitude $m_\lambda$ for each spectral region.  Error bars on these narrowband magnitudes were determined using same method for the unsmoothed spectra.  

\begin{figure*}
    \centering
    \includegraphics[scale = 1]{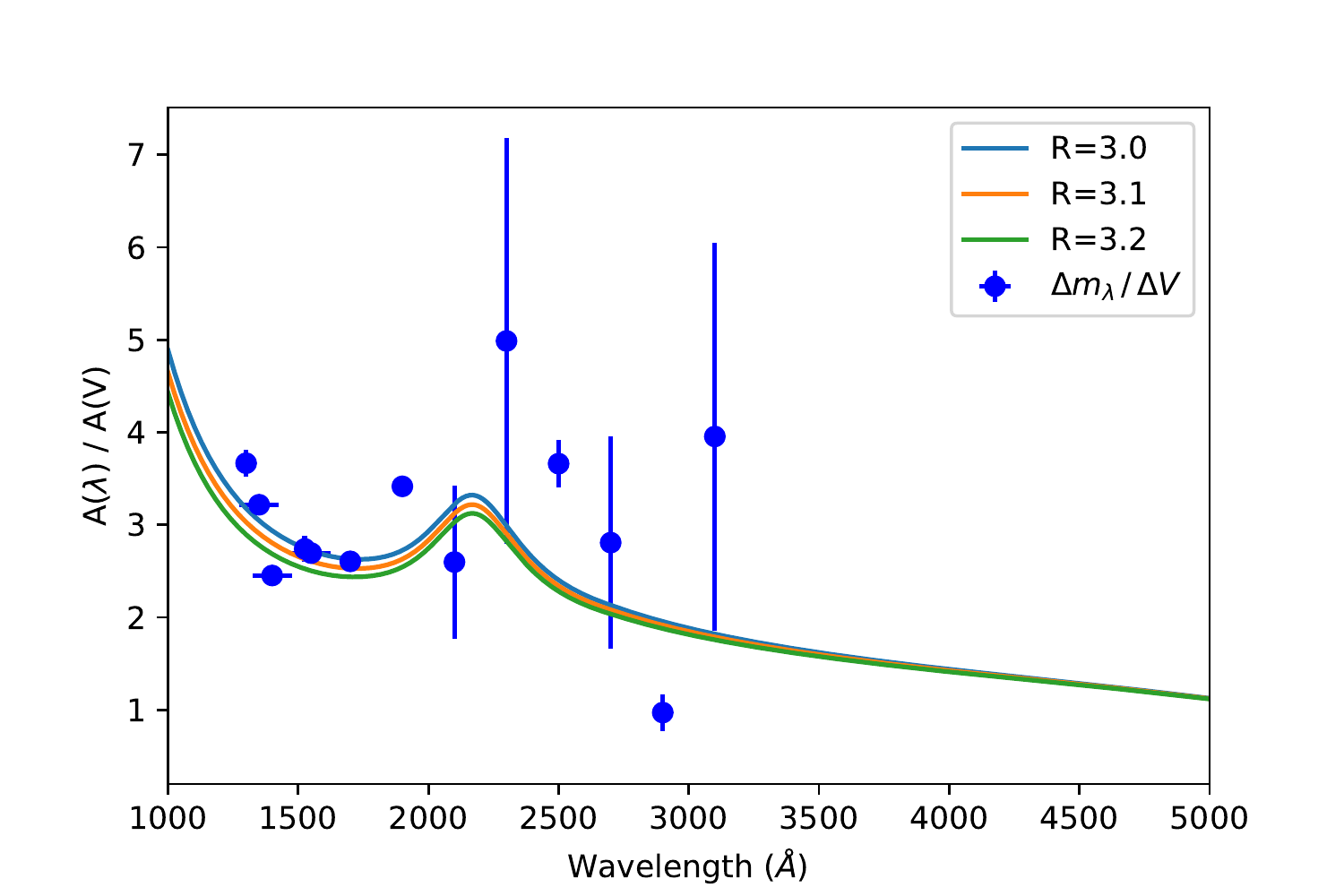}
    \caption{We compare the predicted extinction from F04 to our measured ranges of UV variability.  Here, $\Delta m_\lambda$ is shown relative to $\Delta V$, where $\Delta V$ is scaled to our measured TESS variability, $\Delta m_{\mathrm{TESS}}$, using F04. }
    \label{UVextinction}
\end{figure*}

If optically thin clouds are surrounding the star, then the absorption should vary with rotational phase with an amplitude that is linearly dependent on the number of absorbers.  We use the \texttt{dust extinction} package (an \texttt{astropy} affiliated package) with the F04 reddening model to plot the predicted extinction as a function of wavelength relative to the $V$-band, $A(\lambda) / A(V)$ for several values of the total to selective extinction, $R = A(V) / E(B-V)$.  We then use the F04 model to scale our measured amplitude of TESS variability, $\Delta m_{\mathrm{TESS}}$, to the predicted $V$-band amplitude, $\Delta V$, and we plot our measured UV amplitudes, $\Delta m_\lambda$, relative to $\Delta V$ to compare them to the F04 reddening model in Figure \ref{UVextinction}.  We find that the amplitudes of the UV variability are very consistent with standard interstellar reddening.  

It is unclear whether the UV lightcurves are out of phase with the optical lightcurve maximum, as the $B$-band maximum is, since the uncertainty in the ephemeris of \acvn will translate to very large uncertainties over the many years spread between IUE and TESS missions.  The IUE observations were taken at approximately HJD 2446500 and HJD 2444950, while TESS data are from approximately HJD 2451920.  With such a long time difference, an uncertainty of $\sim 0.001$ d in $T_\mathrm{0}$ translates to an uncertainty of a full rotational period in the time of maxima. The UV lightcurves remain consistent with an in-phase correlation with the TESS lightcurve.

\section{Other Observables from a Possible Magnetosphere in \acvn} \label{other}

\acvn has a surface gravity $\log g = 3.89$ \citep{sikora2019}, and we find its Kepler corotation radius to be $R_\mathrm{K} = 7.5 \; R_\star$.  Assuming a radiatively-driven mass loss rate expected for A-type stars of $\dot{M} = 10^{-12}$  $M_\odot$~yr$^{-1}$ \citep{leto2006} and a wind velocity comparable to the escape velocity of the star ($v_\mathrm{wind} \sim v_{\mathrm{esc}} = 520$ km~s$^{-1}$), such a low mass loss rate would correspond to a wind density $\rho_\mathrm{w} = \dot{M}/(4 \pi v_{\mathrm{wind}} r^2) \approx 8 \times 10^{4}$ particles~cm$^{-3}$ at $R_\mathrm{K}$. 
The corresponding Alfv\'en radius would be $R_\mathrm{A} \approx 35 \; R_\star$ (using our measured $\vtr{B}$ with other parameters from \citealt{shultz2022}).  These values of $R_\mathrm{K} < R_\mathrm{A}$ would place \acvn well within the CM region of a rotation-magnetic wind confinement diagram \citep{petit2013, shultz2022}.

\citet{leto2021}, \citet{shultz2022}, and \citet{owocki2022}  demonstrate convincingly that, among the magnetic hot stars, radio emission from gyrosynchrotron radiation and H$\alpha$ emission are both independent of the stellar mass loss rate, $\dot{M}$.  Instead, they are well correlated to the magnetic field flux and stellar rotation period.  Thus, the latter two papers argue that both the radio and H$\alpha$ emission originate from the same mechanism, a CB.  The accumulation of the outflow into the CM is observed as H$\alpha$ emission.  CB from the outer edge of the CM disk contributes relativistic electrons to the circumstellar environment, leading to the observed radio emission.

There remain several potential problems with the CB model, however.  The first is the lack of optical or X-ray flaring associated with the class of magnetic hot stars.  While \citet{shultz2020} and \citet{owocki2020} attribute this to a relatively steady state, continuous leakage from the CM rather than occasional, violent breakout events, this remains inconclusive.  
An A0 V star like \acvn with a mass loss rate on the order of $10^{-12}$ $M_\odot$~yr$^{-1}$ would presumably fill a magnetosphere containing $10^{-11}$ $M_\odot$ (as we found above) in a timescale, $\tau$, of just a few years.  The magnetosphere essentially acts as a reservoir for the outflow, leaking only once it reaches a critical density.  But in order to maintain conservation of mass, the leakage rate must balance the fill rate, $\dot{M}$, to maintain a steady state CB.   This would clearly contradict the observed lack of relationship between $\dot{M}$ and the gyrosynchrotron emission from the leaking high energy electrons.  
If we remove the requirement that the CB must be steady state, then the reservoir might be emptied in quasi-periodic, short-lived bursts that instead depend on the product $\dot{M} \times \tau$, but these would likely be observed as high energy flaring events.  The filling timescale is of the order $\tau \sim 10^0$ to $10^2$ years for the class of \acvn stars, so flaring events should be commonly observed among the 1169 \acvn variables catalogued in Simbad.  Such a CB flare has been observed only once in the Ap star CU Vir \citep{das2021}.  $\sigma$ Ori E also has $\tau \sim 200$ years for its CM \citep{townsend_owocki2005}.  For hotter Bp stars with H$\alpha$ emission, discrete bursts would also cause a largescale disruption or the entire disappearance of H$\alpha$ on the timescale $\tau$.  As neither flares nor H$\alpha$ disappearance have been observed for nearly all Ap or Bp stars, we conclude that discrete CB events also poses a contradiction.

\acvn also does not fit well into the CB paradigm because it is radio bright \citep{becker1995, drake2006, leto2021} yet lacks H$\alpha$ emission or even shell absorption.  We reviewed more than 500 optical spectra of the star in the PolarBase database to confirm this for ourselves.   \citet{owocki2020} propose that this class of stars may still experience continuous CM leakage, but without a large reservoir to produce H$\alpha$ emission.  Yet the conservation of mass in this proposed model still contradicts the observation that radio brightness of H$\alpha$-faint systems is independent of their mass loss rates.
There remain several other reasons that \acvn might have a source of relativistic electrons but not exhibit H$\alpha$ emission in this scenario.
Could its circumstellar environment simply be too cool to have a significant population of excited H atoms to participate in emission?  The entire class of cooler Bp and Ap stars does not exhibit H$\alpha$ emission as a rule  \citep{owocki2020}.  However, classical Be star emission is observed across the full range of B spectral types and even down into the A0 spectral class, as listed in the Be Star Spectra Database \citep{neiner2011}.  Pre main-sequence Herbig Ae/Be stars also exhibit somewhat similar emission properties (e.g. \citealt{vioque2020}). Clearly the circumstellar environments of early A-type stars are sufficiently hot.

It has also been proposed that the mass outflow is dominated by a metal-ion wind in cooler Bp and Ap stars, filling the CM with metals but not H \citep{owocki2020}.  This would provide a feasible explanation for the missing H$\alpha$ emission while still allowing for the presence of free electrons.  However, a rigid, metal-rich CM disk should be expected to cause irregular continuum photometric variability more similar to $\sigma$ Ori E and other CM-host stars. 

Instead of a CM, we can consider the consequences of a putative DM on the H-alpha emission and other observable properties of \acvn.
Our DM model for \acvn clearly implies an outflow that is about 7 orders of magnitude higher than expected for a radiatively-driven wind, implying $\dot{M} \sim 10^{-5}$ $M_\odot$~yr$^{-1}$.  At our measured $|\vec{B}| \sim 0.8$ kG, the magnetic energy density is $u_\mathrm{B} = B^2/8\pi = 2500$ N~m$^{-2}$, which is three orders of magnitude greater than the radiation pressure $P_{\mathrm{rad}} = \frac{4}{3c} \sigma T^4 = 2.5$ N~m$^{-2}$.  Clearly radiation pressure is not sufficient to drive such an outflow, but there is ample energy from the magnetic field itself.  After all, our Sun does not have a radiatively-driven outflow either; its magnetic energy heats the corona, increasing the thermal pressure gradient that drives the solar wind \citep{lamers1999}. \citet{Giarrusso_2022} notes that a magnetic gradient pumping mechanism could drive mass loss into the magnetospheres of Bp stars.
So, a high mass loss rate could be physically possible in these low luminosity Ap stars.

\begin{figure}
    \centering
    \includegraphics[scale = .3]{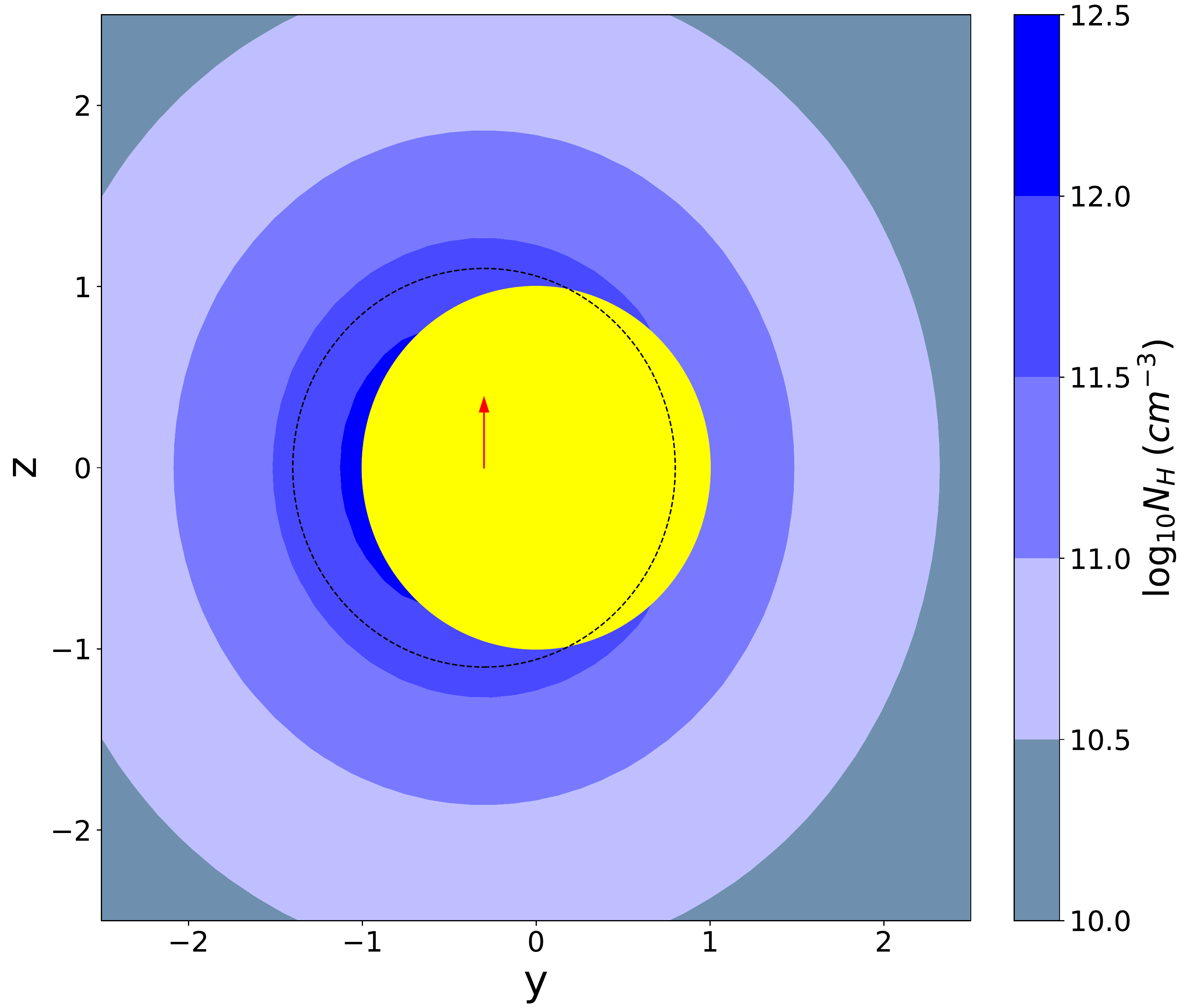}
    \includegraphics[scale = .3]{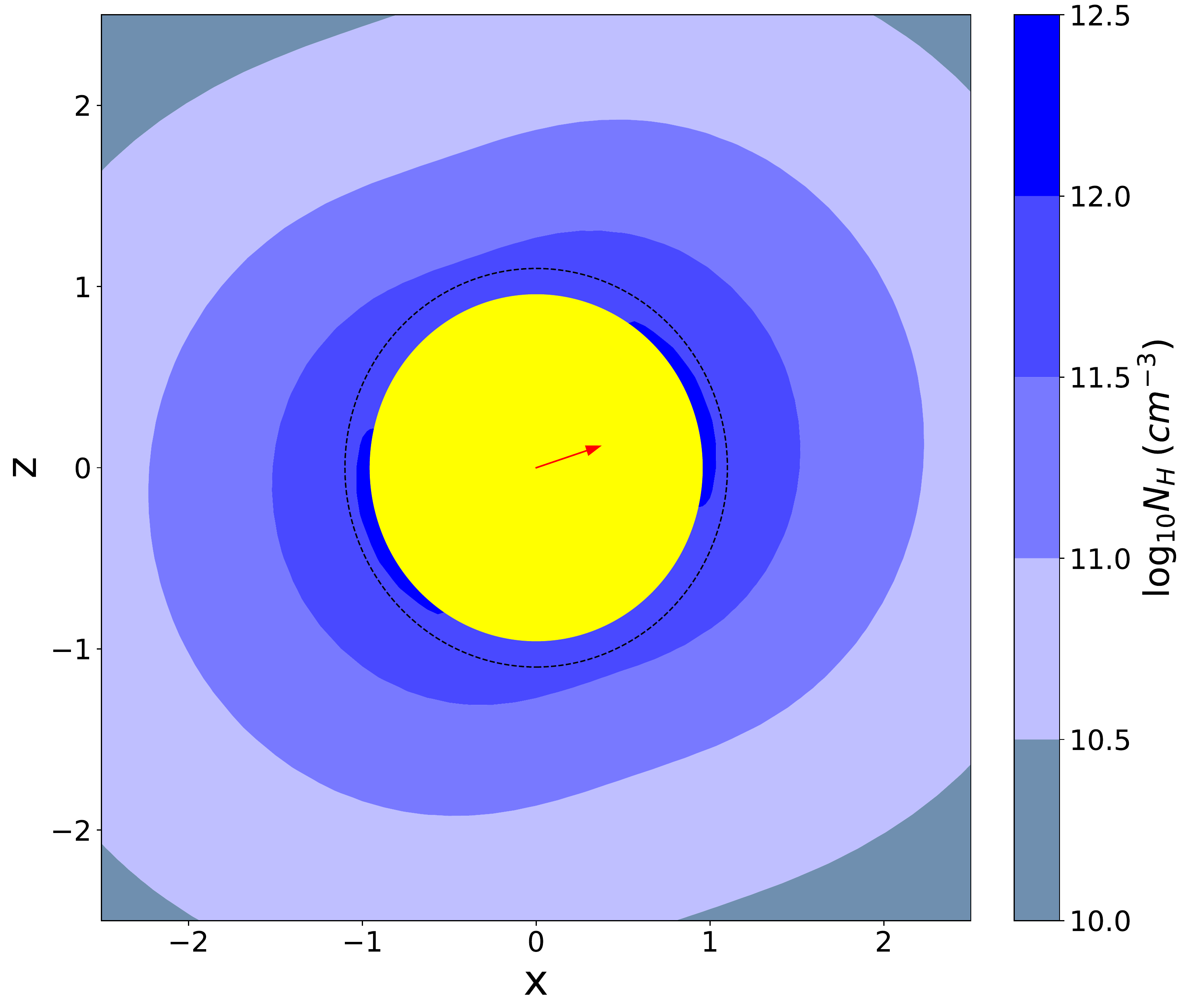}
    \caption{We show two slices of the DM around the star with particle densities indicated by the contours. The densities were calculated using our best fit model parameters of $(x_{\mathrm{0}},y_{\mathrm{0}},z_{\mathrm{0}},\beta) = (0, -.3R_\star, 0, 4 \pi/10)$. The left figure shows a slice in $y$ and $z$ located at $x = x_{\mathrm{0}} = 0$. The right figure shows a cross section in $x$ and $z$ located at $y = y_{\mathrm{0}} = -.3R_\star$. An Alfv\'en radius of $1.1R_\star$ is shown as a black dashed line, which we center on the magnetic field origin, and the dipole moment axis is indicated by the red arrow. The yellow circle indicates the radius of the star; it is slightly smaller on the right figure as the slice is taken at $y = y_{\mathrm{0}} = -.3R_\star$ }
    \label{CrossSection}
\end{figure}

With the higher $\dot{M}$ calculated from our DM model, the corresponding Alfv\'en radius reaches nearly down to the stellar surface, $R_\mathrm{A} \approx 1.1 R_\star$ using the prescription of \citet{ud_Doula_2002}.  But of course, this simplified order-of-magnitude estimate neglects the asymmetry of the field and assumes isotropic mass loss, so the true Alfv\'en surface is likely far more complex.  We show in Figure \ref{CrossSection} the contours of the particle density of our proposed DM as viewed over two different slices through the star.  We find that the circumstellar material would be highly confined very close to the stellar surface, consistent with a small value of $R_\mathrm{A}$. 

Ultimately, though, the peak density of our modeled DM in \acvn should be comparable to the densities observed in $\sigma$ Ori E \citep{townsend2013, owocki2020}, where H$\alpha$ emission is clearly present and modulated over the stellar rotational period.  The lack of H$\alpha$ emission or shell absorption at every rotational phase in \acvn forces us to conclude that a DM is unlikely to be present.

Furthermore, the high mass loss rate required for a DM in \acvn would be expected to produce a very large radio photosphere with bright thermal radio emission \citep{lamers1999,  kurapati2017}.  While some radio emission has been observed from \acvn, it a few orders of magnitude fainter than expected for such a windy system.

\section{Conclusions}

We have modeled the star \acvn using a rather unconventional simplified DM to test whether this star may harbor a physical magnetosphere.  Our tilted, offset magnetic dipole does a very good job of reproducing the \textit{TESS} lightcurve and the surface magnetic field as observed by \citet{silvester_kochukhov_wade_2014}.  The UV variability, as observed from \textit{IUE} spectra, is consistent with traditional reddening models but is not conclusive.  On the other hand, the strong stellar outflow required to produce our modeled DM should be observable via H$\alpha$ emission and bright, thermal radio emission, neither of which has been observed.  We have also shown that a CB mechanism is unlikely to be present in \acvn and its counterparts.  We conclude that the chemical spot model remains the best explanation for the peculiar variability of \acvn, but questions still remain about the origin of its non-thermal radio emission.  This continues to be an interesting source worth further investigation.

\section{Acknowledgements}

We are extremely grateful to the referee for their comments that significantly improved this paper.  We are also grateful to James Silvester and Oleg Kochukhov for sharing their magnetic field measurements of \acvn and for helpful discussions about this project.  We also thank Joshua Pepper for further discussions about the TESS mission data, and we thank Thomas Cahill and Lehigh University for their generous support for this work. 
This paper includes data collected by the TESS mission, which are publicly available from the Mikulski Archive for Space Telescopes (MAST). Funding for the TESS mission is provided by the NASA's Science Mission Directorate.
This research has made use of the SIMBAD database, operated at CDS, Strasbourg, France.
This research uses services or data provided by the Astro Data Lab at NSF's National Optical-Infrared Astronomy Research Laboratory. NOIRLab is operated by the Association of Universities for Research in Astronomy (AURA), Inc. under a cooperative agreement with the National Science Foundation.

\software{astropy \citep{astropy2018}, astroquery \citep{astroquery2019},  Jupyter \citep{jupyter2016}, matplotlib \citep{matplotlib2007}, numpy \citep{numpy2011}, pandas \citep{pandas2010}, scipy \citep{scipy2020} } 

\bibliographystyle{aasjournal}
\bibliography{bib}

\end{document}